# Lattice dynamics across the ferroelastic phase transition in $Ba_2ZnTeO_6$: A Raman and first-principles study

Shalini Badola[1], Supratik Mukherjee[2], B. Ghosh[1], Greeshma Sunil[1], G. Vaitheeswaran[3], A. C. Garcia-Castro[4], and Surajit Saha[1]*

[1]*Indian Institute of Science Education and Research Bhopal, Bhopal 462066, India*

[2]*Advanced Center of Research in High Energy Materials (ACRHEM), University of Hyderabad, Prof. C. R. Rao Road, Gachibowli, Hyderabad 500046, Telangana, India*

[3]*School of Physics, University of Hyderabad, Prof. C. R. Rao Road, Gachibowli, Hyderabad 500046, Telangana, India*

[4]*School of Physics, Universidad Industrial de Santander, Calle 09 Carrera 27, Bucaramanga, Santander, 680002, Colombia*

*\*Correspondence: surajit@iiserb.ac.in*

**ABSTRACT**: Structural phase transitions drive several unconventional phenomena including some illustrious ferroic attributes which are relevant for technological advancements. With this note, we have investigated the structural transition of perovskite-type trigonal $Ba_2ZnTeO_6$, across $T_c \sim 150$ K, which is also accompanied by a para- to ferroelastic transition. With the help of Raman spectroscopy and density-functional theory (DFT)-based calculations, here we report new intriguing observations associated with the phase transition in $Ba_2ZnTeO_6$ elucidating the lattice dynamics across the $T_c$. We have observed the presence of a central peak (quasi-elastic Rayleigh profile), huge softening in the soft mode, hysteretic phonon behavior, and signatures of coexistent phases. The existence of a central peak in $Ba_2ZnTeO_6$ is manifested by a sharp rise in the intensity of the Rayleigh profile in concomitant with the damping of the soft mode near $T_c$, shedding light on the lattice dynamics during the phase transition. While most of the phonon bands split below $T_c$ confirming the phase transition, we have observed thermal hysteretic behavior of phonon modes that signifies the first-order nature of the transition and presence of coexisting phases, which are corroborated by our temperature-dependent x-ray diffraction and specific heat measurements.

Further, an evidence of the concomitant structural transition appears in the form of huge softening in the thermal response of the soft phonon mode at ~ 31 cm$^{-1}$ which is remarkable compared to the hitherto known behavior of soft modes in well-known ferroelectrics. This is further corroborated by our phonon calculations that show an unstable $E_g$-mode in the high-symmetry structure involving TeO$_6$ octahedral rotation (with Ba and Zn translation) which later condenses into the *C2/m* low-symmetry phase.

## 1. INTRODUCTION

The investigation of phase transition witnessed a surging interest in the last few decades not only from a technological but fundamental viewpoint as well. Structural phase transitions possessing strong coupling to the magnetic, electric, and elastic degrees of freedom are of immense importance as they demonstrate remarkable effects like magneto-structural transition[1,2], metal-insulator transition[3,4], and sometimes, ferroic orders as well[5]. While ferroic orders like ferroelectricity and ferromagnetism are frequently encountered, the occurrence of ferroelasticity is comparatively rare and restricted to a selected class of structures that follow certain symmetry rules upon transition. Notably, Aizu listed down 94 active ferroelastic species that facilitate the identification of ferroelastic crystals based on lowering of the point group symmetries[6,7]. The phenomenon of ferroelasticity involves a structural transition from high- to low-symmetry phase along with lattice distortion[6,7]. As a consequence, a significant impact can be expected on the phonon (lattice) properties. Of particular interest in this context are the low-frequency phonons that shed light on the dynamics of the lattice where soft modes and lattice relaxation processes play vital roles in the phase transition.

The concept of "soft phonon" or "soft mode" offers a powerful tool to establish a fair understanding of the mechanisms (displacive and/or order-disorder) responsible for a phase transition. Reportedly, the involvement of underdamped or overdamped phonons entitles such classification, as noted in the case of PbTiO$_3$ and BaTiO$_3$[8-10]. Crucial details of the nature (order) of the phase transition, associated crystal symmetries, and underlying lattice instabilities can also be understood from soft phonons[11, 12]. Basically, soft phonons are unstable normal modes of lattice that exhibit a pronounced decrement in phonon frequency leading to the vanishing of the mode across the phase transition temperature (T$_c$). In addition, the soft phonons drive a change in the crystal symmetry under the effect of temperature which is dictated by the eigenvectors (atomic displacements) of the modes and their associated anharmonicities[11]. A pronounced anharmonicity in the soft mode may arise due to lattice instabilities that become prominent near T$_c$. Some of the examples of the classic perovskite systems where a strong anharmonicity, which may be estimated as $\frac{\Delta\omega}{\omega} = \frac{\omega(T) - \omega(T_c)}{\omega(T_c)}$ % when the quasi-harmonic contribution is negligible, is associated with the soft phonon driving a phase transition are SrTiO$_3$[13], KNbO$_3$[14], and SrZrO$_3$[15]. It is interesting to note that the soft mode-driven phase transition in ferroic systems is often accompanied by dynamical effects, which are manifested as a central peak (quasi-elastic Rayleigh peak) thus shedding light on the evolution (dynamics) of the lattice during the phase transition. The central peak originates from the phonon density and entropy fluctuations[16, 17], lattice defects[18], ion displacements[19], domain wall motion[20] and, therefore, it needs a careful and detailed investigation to elucidate the dynamics and the association with the soft mode.

Notably, most of the systems cited above belong to the usual perovskite structure (cubic/tetragonal/orthorhombic phases) at ambient conditions, with a tolerance factor ($t = \frac{r_A + r_O}{\sqrt{2}(r_B + r_O)}$, where $r_A$ and $r_B$ represent the cationic and $r_O$ represents the anionic radii) close to unity. For $t > 1$, hexagonal variants of perovskites get stabilized possessing large extended unit cells with strong structural anisotropy (large *c/a* ratio). As a result, the phonon anharmonicity in the hexagonal isomorphs can be really remarkable as compared to their cubic or tetragonal counterparts. One such variant of the double-perovskite family with $t > 1$ is $Ba_2ZnTeO_6$ (BZT) which crystallizes into a trigonal symmetry (R-3m space group) at ambient conditions[21] and cubic phase when synthesized at high pressure (~ 4GPa)[22]. Moreira *et al.* have shown that BZT undergoes a structural phase transition from trigonal to monoclinic phase below 140 K, satisfying the symmetry requirements for ferroelasticity[23]. On the other hand, an isostructural system $Ba_2NiTeO_6$ lacks such a transition down to 1.8 K[24]. Such tellurium-based double-perovskite type systems have also been proposed to exhibit strong anharmonicity due to the rattling motion of Te atoms in large octahedral framework[25]. These studies indicate the strong relevance of internal lattice pressure and anharmonicity in determining the physical properties of BZT and similar systems.

Inelastic light scattering is a robust and simple technique to probe phonon dynamics. Here, with a systematic temperature-dependent Raman investigation, we present some remarkable aspects of the phase transition in $Ba_2ZnTeO_6$ which are crucial to elucidate the lattice dynamics across the transition. We have discussed (a) the signatures of the 'central peak' near $T_c$ indicating the involvement of lattice relaxation processes, (b) the hysteretic behavior of phonon modes across the phase transition, thus, identifying it to be of first-order in nature as well as the presence of coexistent high temperature trigonal and low-temperature monoclinic phases in the hysteretic region, (c) a very strong anharmonicity of the soft phonon mode (at ~ 31 cm$^{-1}$) in comparison with commonly studied perovskite-based ferroelectrics like $BaTiO_3$ and $PbTiO_3$ and (d) about the involvement of $TeO_6$ octahedra along with Ba and Zn translation driving the phase transition. Further, we have quantitatively analyzed the quasi-harmonic and anharmonic contributions to invoke the role of anharmonicity in the observed phase transition. We have analyzed these observations with the lowering of the crystallographic symmetry across the phase transition and corroborate our results with temperature-dependent specific heat and x-ray diffraction measurements. Our density functional theory (DFT)-based calculations show that the rotations of $TeO_6$ octahedra along with Ba and Zn translation associated with the soft mode are responsible for the structural transition to the monoclinic phase and support the presence of a strong anharmonicity.

## 2. SYNTHESIS, EXPERIMENTAL, AND COMPUTATIONAL DETAILS

Solid state reaction method was used to synthesize polycrystalline $Ba_2ZnTeO_6$ with high purity (99.995%) oxide precursors such as $BaCO_3$, $ZnO$, and $TeO_2$ obtained from Sigma Aldrich. These precursors were mixed in stoichiometric ratio and later treated sequentially at 750°C, 900°C, and 1050°C for 4, 6, and 4 hours, respectively. The phase purity of as-synthesized powder was confirmed using room temperature x-ray diffraction (XRD) employing PANalytical diffractometer with Cu-$K\alpha$ radiation ($\lambda = 1.5406$ Å). A Rietveld phase analysis performed using Highscore Plus software detected a minor fraction of $Ba_3TeO_6$ (~ 2.1 %) as secondary phase besides the trigonal phase of $Ba_2ZnTeO_6$ (97.9 %). A compositional analysis of the BZT was obtained using the Energy Dispersive X-Ray spectrometer (EDAX) of a High-Resolution field emission scanning electron microscope (HR-FESEM) from Zeiss (Model: ULTRA plus). Further, the structural evolution of the BZT was studied using temperature-dependent XRD employing Anton Paar 450 low-temperature stage from 90 K up to 350 K. To probe the crystal structure and lattice dynamics further in detail, temperature-dependent phonon measurement was performed in the range of 80 to 400 K using LabRAM-HR Evolution Raman spectrometer combined with Linkam heating stage (Model: HFS600E-PB4) and Peltier-cooled charge coupled device (CCD) detector. Raman spectra were acquired in the backscattering geometry with a laser excitation source of wavelength 532 nm (Nd:YAG). The phase transition was further confirmed by measuring specific heat in the temperature range of 80 to 200 K using Physical Property Measurement System (PPMS Re-Liquefier, Quantum Design).

We performed DFT[26, 27] calculations as implemented in the VASP code (version 5.4.4)[28, 29]. The projected-augmented waves (PAW)[30], approach was used to represent the valence and core electrons. The electronic configurations considered in the pseudo-potentials for the calculations were Ba:($5s^25p^66s^2$, version 06Sep2000), Zn:($4s^23d^{10}$ version 06Sep2000), Te: ($5s^25p^4$ version 08Apr2002), and O: ($2s^22p^4$, version 08Apr2002). The exchange-correlation was represented within the generalized gradient approximation GGA-PBEsol parametrization[31]. The periodic solution of the crystal was represented by using Bloch states with a Monkhorst-Pack[32] $k$-point mesh of $7\times7\times7$ and 600 eV energy cut-off to give forces convergence within the error smaller than 0.001 eV·Å$^{-1}$. Born effective charges and phonon calculations were performed within the density functional perturbation theory (DFPT) approach[33] as implemented in the VASP code. Phonon dispersions were post-processed in the Phonopy code[34]. The atomic structure figures were elaborated with the support of the VESTA (Visualization of Electronic and STructural Analysis) code[35].

## 3. RESULTS AND DISCUSSIONS

### 3.1 Phase determination and phonon assignments at Room-temperature

At room temperature, BZT crystallizes in the trigonal R-3m symmetry. As presented in Fig. 1(a), the XRD pattern of BZT at room temperature is refined using the crystallographic information file no.: 25005 of the isostructural system Ba$_2$NiTeO$_6$ (BNT) adapted from the inorganic crystallographic structural database (ICSD)[36]. The lattice parameters of the trigonal phase obtained in the hexagonal representation from the Rietveld refinement of the XRD profile were found to be $a_t = b_t = 5.8236$ Å and $c_t = 28.692$ Å at room temperature, which are in agreement with earlier reports[22, 23]. The trigonal crystal structure of BZT comprises of corner- and face-shared $ZnO_6$ and $TeO_6$ octahedral units as shown in Fig. 1(b), giving rise to a unit cell of 12-layer (12R) type thus, possessing a strong structural anisotropy (large $c/a$-ratio). The compositional (elemental) analysis of the BZT, shown in Fig. 1(c), was performed using EDAX measurements that suggest the cations to be in nearly 2: 1: 1 ratio which is in reasonable agreement with the expected stoichiometry (Ba: Zn: Te: O = 2: 1: 1: 6). Since the EDAX measurement cannot accurately estimate the elements with lower atomic numbers (Z < 10), the oxygen stoichiometry cannot be reliable. Further details of the structure and crystal symmetry of BZT were obtained by Raman spectroscopic measurements which are discussed below.

Figure 2 shows the Raman spectrum of BZT at room temperature which is deconvoluted into 15 phonon modes (labeled as Z1 to Z15) by fitting with the Lorentzian function. The factor group analysis predicts a total of 16 Raman active phonons for the R-3m space group (trigonal structure) of BZT that comprise of 7 $A_{1g}$ and 9 $E_g$ symmetric modes (along with 2 $A_{2g}$ silent modes Z1′ and Z6′ listed in Table I). We, however, could clearly detect 6 $A_{1g}$ and 9 $E_g$ phonons in our measurements. A peak marked with asterisk (*) in Fig. 2 (at around ~ 135 cm$^{-1}$) could possibly be the seventh $A_{1g}$ mode which is weak in intensity and, therefore, its thermal behavior could not be studied. Here, it is to be noted that our room temperature Raman spectrum of BZT matches to a great extent with the earlier report[23]. However, a few dissimilarities were identified between the previous report and our data upon comparison. Therefore, we have revisited the assignments of the modes based on group-theoretical analysis[37], our Raman measurements, DFT calculations and the available polarization-dependent Raman data from the previous report[23]. A detailed account on the same along with the phonon eigenvectors is provided in the supplementary information. Importantly, our calculation could show an unstable $A_{2g}$ mode (Z1′) in the trigonal phase which couples with the soft mode (Z1) (discussed later). The modes Z9 and Z15 could not be assigned theoretically in the high-temperature phase while the modes Z8 and Z15 in the low-temperature phase could not be theoretically assigned. Symmetry assignments of the modes are listed in Table I.

### 3.2 Temperature dependence of phonons

Raman measurements were performed as a function of temperature in the range of 80 – 300 K in the heating and cooling cycles (and in the extended temperature range of 80 to 400 K in the heating cycle, shown in supplementary information). The measurements were done with a step size of 10 K (with a rate of 15°C/min) and a waiting period of 30 minutes to attain stable sample temperature before acquiring the spectrum. As several of the modes split below ~150 K, the Raman spectrum was resolved with 25 Lorentzian peaks at lower temperatures. The increase in the number of modes at lower temperatures is an indicative of lowering of the crystal symmetry (a structural transition) associated with the para- to ferroelastic transition at ~ 150 K[23]. Figure 3 presents the evolution of the Raman bands with temperature from which it is evident that 9 phonon bands (Z1, Z3, Z5, Z6, Z7, Z8, Z11, Z12, and Z15) split below the transition temperature lifting the degeneracy of the $E_g$ modes into $A_g$ and $B_g$ symmetries in the low symmetry monoclinic phase. Further, Z2 and Z4 show an unusual blue-shift in phonon frequency upon cooling below 150 K as discussed later. Figure 4 presents the temperature response of frequencies of all the phonon modes of BZT. While the thermal response of most of the phonons is in agreement with the earlier report[23], we notice a disagreement in the symmetry assignment based on the thermal behavior of some of the modes like Z3, Z6, Z9 and Z13, as discussed in the supplementary information. Since some of the phonons exhibit unusually large shifts in frequency over temperature besides splitting of the bands, it is important to understand the origin of such large shifts and their effects on the ground state properties, which are discussed below.

### 3.3 Signatures of the central peak and strong anharmonicity

The variation of phonon frequency as a function of temperature is usually described as[38, 39]

$$\omega(T) = \omega(0) + \Delta\omega_{qh}(T) + \Delta\omega_{anh}(T) + \Delta\omega_{el-ph}(T) + \Delta\omega_{sp-ph}(T). \qquad (1)$$

where, $\omega(0)$ is the phonon frequency at 0 K while the quasi-harmonic term, $\Delta\omega_{qh}(T)$, represents the volume contribution arising from the thermal expansion of the lattice. The term $\Delta\omega_{anh}(T)$ refers to the change in frequency due to intrinsic anharmonic correction in the self-energy arising from phonon-phonon interactions. On the other hand, $\Delta\omega_{sp-ph}(T)$ is the contribution due to spin-phonon interaction renormalizing the phonon frequencies, which can be neglected in BZT due to its non-magnetic nature. The term, $\Delta\omega_{el-ph}(T)$, takes into account the renormalization in phonon frequency due to electron-phonon interaction, which is also absent in the case of electrically insulating BZT. Therefore, it can be inferred that the major renormalization of the phonon frequencies upon varying temperature in BZT occurs mainly due to the contribution from the quasi-harmonic (lattice expansion) and the intrinsic phonon-phonon anharmonic interactions. On analyzing the volume change of the trigonal phase from XRD data (refer to supplementary information for details), we infer that the major contribution to the unusual shift in the soft mode (Z1) frequency is rendered by the phonon-phonon anharmonic interactions (originated from lattice instabilities) as compared to a negligibly small quasi-harmonic contribution.

The temperature dependence of mode Z1 suggests it to be a soft mode (see Figs. 3 and 4).

The manifestation of soft modes can be noticed in various ways through the thermal response of phonons. Typically, the frequency of a soft mode tends to zero as the temperature approaches $T_c$ from above exhibiting anomalous behavior (i.e. a decrease in frequency with decreasing temperature) whereas it shows a normal trend (i.e. increase in frequency with decreasing temperature) below $T_c$. In general, the behavior of soft modes as a function of temperature can be captured by Cochran's relation which suggests that[40, 41]

$$\omega(T) = A\,(T - T_c)^{1/2}, \qquad (2)$$

where, $\omega(T)$ denotes the temperature-dependent frequency, $T_c$ refers to the phase transition temperature, while $A$ represents the fitting parameter. The behavior suggested for $\omega(T)$ in Equation 2 is generally applicable to displacive-type phase transition[11, 42]. The frequency response of the Z1 mode agrees well with Cochran's relation (Eq. 2) (for $T_c \sim 150$ K), shown in Fig. 4, suggesting the displacive-nature of the phase transition in BZT, as also reported earlier[23]. Notably, an underdamped soft phonon leads to the displacive-nature of a phase transition[8, 43]. Since the frequency ($\omega$) of the soft mode (Z1) is larger than its linewidth ($\Gamma$) in the entire temperature range (see Fig. 5(a)), the phonon mode can be interpreted as underdamped which further suggests a dominant displacive-type mechanism for the phase transition in BZT. However, it should be noted that the damping of the soft mode increases close to the transition temperature (with $\frac{\Gamma}{\omega}$ tending to unity) which may lead to the presence of a central peak (Rayleigh quasi-elastic scattering profile) as discussed later. It is worth mentioning here that such an approach to identify the dominant displacive-type mechanism of the phase transition was not established before in BZT. Our DFT-based calculations, discussed below, further confirm the displacive nature of the transition. More importantly, the mode Z1 splits below the transition and the shift in frequency over temperature is unusually large in both the phases across the transition. However, a complete softening (i.e. phonon frequency going to zero) could not be observed for Z1 and the mode disappeared (near 170 K upon cooling) before the transition, suggesting it to be first-order in nature[11]. The Raman spectra recorded in the extended temperature range of 80 - 400 K (see Fig. S1 in supplementary information) were employed to estimate the phonon anharmonicity in the soft mode. We have estimated the change in phonon frequency and anharmonicity using the relation $\frac{\omega(400\ K) - \omega(170\ K)}{\omega(170\ K)}$. It is to be noted that we have considered the phonon frequency at 170 K (instead of $T_c \sim 150$ K) for the estimation of anharmonicity as the phonon mode is not resolvable (visible) at temperatures close to $T_c$ due to a concomitant increase in the intensity of the Rayleigh profile at the vicinity of $T_c$ (discussed later). Strikingly, for the Z1 mode, the high-temperature phonon branch with $E_g$ symmetry shows a change in frequency by ~ 120 % upon cooling from 400 K to 170 K (refer supplementary information for more details). Further, Z1 mode ($E_g$ symmetry) splits into two phonons with $A_g$ and $B_g$ symmetry at lower temperatures below the structural transition. The $B_g$ phonon branch is found to demonstrate another unusually large shift of about 145 % below $T_c$ down to 80 K. In absence of any significant quasi-harmonic contribution, these unusually large phonon shifts of the soft mode can be attributed to a very strong anharmonicity. A quantitative analysis of the quasi-harmonic and anharmonic contributions to the phonon shifts is provided in the supplementary information[44]. To contemplate the origin of such a strong anharmonicity that leads to unusually large frequency shifts in BZT, we have compared its anharmonicity ($\frac{\Delta\omega}{\omega}$ % for the soft mode) with some of the classic ferroelectric perovskite systems like $BaTiO_3$ and $PbTiO_3$. As shown in Fig. 5(b), BZT demonstrates anharmonicity comparable to $BaTiO_3$[45] but lesser than that of $PbTiO_3$[46]. It should be particularly noted that the temperature ranges (i.e. temperature of the sample) in which the $\frac{\Delta\omega}{\omega}$ % values are estimated for PbTiO$_3$ and BaTiO$_3$ are far higher (than that for BZT) where the anharmonic effects are supposed to be more pronounced owing to higher temperatures. Moreover, PbTiO$_3$ exhibits a negative thermal expansion coefficient from room temperature to ~ 763 K[47]. Therefore, one may expect a large anharmonicity for such (i.e. BaTiO$_3$ and PbTiO$_3$) systems. On the contrary, the estimated anharmonicity for BZT exists even at much lower temperatures with a positive thermal expansion coefficient as observed in our temperature-dependent x-ray diffraction data (refer Fig. S3 of supplementary information). Therefore, we believe that the unusually large value of $\frac{\Delta\omega}{\omega}$ % (anharmonicity) in BZT originates from the lattice instabilities (in the soft mode Z1) during the ferroelastic phase transition, giving rise to TeO$_6$ octahedral rotations with Ba and Zn displacements, as also suggested from our DFT calculations (discussed later). Notably, the large frequency-shift is seen not only for the mode Z1 but for the other modes as well. Namely, the modes Z2 and Z4 display an increase in their phonon frequencies by ~ 3 % and 4 %, respectively, below the phase transition (between 150 – 80 K). These unusual shifts in Z2 and Z4 below $T_c$ may be attributed to lattice renormalization at low temperatures below $T_c$ and associated anharmonicities.

Our Raman investigation of the low-frequency region of BZT further shows an important and new observation. It is observed that as we approach the transition temperature, the intensity of the central peak, that is the Rayleigh profile, grows rapidly and then falls off again away from the $T_c$, as shown in Fig. 5(c). Simultaneously, the intensity of the soft phonon band (see inset of Fig. 5(c)) weakens close to the $T_c$ suggesting a possible coupling between the two modes (i.e. the central peak and the soft phonon) that results in the spectral weight (intensity) transfer between the modes[48]. This is an important observation that resembles closely with the reported behavior and phonon dynamics in Pb$_5$Ge$_3$O$_{11}$[49], Li$_2$RuO$_3$[50], AgNbO$_3$[51], and Gd$_2$(MoO$_3$)$_4$[48] in the low-frequency region. Such a behavior of the central peak is often connected with the internal relaxation processes of the crystal which are activated due to an interaction with the dynamical lattice defects[18], phonon density as well as entropy fluctuations[16, 17], and motion of the domain walls[20]. This is evident from the behavior of the linewidth of the central peak (see Fig. 5(d)) which decreases as we approach the $T_c$. As shown in the Fig. 5(d), an increase in the relaxation time of the central peak near $T_c$ is an indicative of slowing down of the lattice dynamics which implies the presence of additional relaxation mechanism involving dynamical lattice defects, phonon density, and possible domain wall motion during the transition (refer supplementary information for more details).[50, 52]

### 3.4 Hysteretic phonon behavior and its origin

The thermal response of the mode Z1 is suggestive of a first-order phase transition. Aiming to further verify the nature (order) of the phase transition, Raman spectra were recorded both in the heating as well as cooling cycles to examine the response of the phonons to thermal cycle. Upon careful observation, it can be noted that the modes Z7, Z9, Z10, and Z11 show hysteretic behavior in their temperature-dependent frequencies in the temperature range of 150 K to about 230 K, as shown in Fig.6. Since thermal hysteresis is an inherent property of first-order phase transition[53], it can be inferred that the observed thermal hysteresis in the phonon behavior of BZT is driven by first-order para-to-ferroelastic phase transition. Moreover, the appearance of thermal hysteresis with temperature may be associated with the co-existence of both the phases (due to nucleation of one phase in the other) in this temperature range. Therefore, we believe that the monoclinic phase could coexists with the trigonal phase in the temperature range 150 K < T < 230 K thus leading to the observed thermal hysteresis, as also evidenced by XRD measurements (discussed in the next section). The thermally induced motion of the domain walls in this coexisting phase likely contributes to the central peak discussed above. Notably, similar hysteretic effects in phonon modes were reported by Perry *et al.* in BaTiO$_3$[54] across the structural transition, though in a narrower temperature range as compared to BZT.

### 3.5 Specific heat and temperature-dependent X-Ray Diffraction

Our Raman data, discussed above, confirm the structural transition of BZT to be first-order in nature. In order to gain more insights of the transition, specific heat measurements were carried out in the temperature range of 80 to 200 K. We observe a broad anomaly in the specific heat near 150 K, as shown in Fig. 7(a), which is an indicative of the evolution of the lattice dynamics around that temperature range where phases coexist, as discussed above. The observed anomaly in specific heat is a manifestation of the phase transition in BZT, thus corroborating our Raman data. The total specific heat ($C_p$) can be represented as a sum of the lattice ($C_{latt}$), magnetic ($C_{mag}$), and electronic specific heat ($C_{el}$) i.e. $C_p = C_{latt} + C_{mag} + C_{el}$[55, 56]. In the present case, the insulating and non-magnetic behavior of BZT eliminates the contribution from the electronic and magnetic terms, respectively. As a result, the total specific heat can be ascribed to the lattice contributions. To be noted that we do not observe a sharp feature in the specific heat at the transition but a smeared broad peak feature which is likely due to the coexistence of both the low and high temperature phases near the $T_c$ (additional details are provided in the supplementary information), thus further corroborating the insight given by our Raman data discussed above.

To gain further clarity, we have performed temperature-dependent x-ray diffraction measurements while heating to examine the changes occurring in the crystallographic structure around the transition while heating. Figure 7(b) displays the evolution of the XRD patterns of BZT at a few typical temperatures that reveal the presence of additional reflections at 2θ = 22.7º and 24.2º in the low temperature regime (below ~ 230 K) which is a clear evidence of the structural transition from a low symmetry monoclinic phase (at low temperature) to the high symmetry trigonal phase (at high temperature). We observe that the new reflection at 24.2°, inherent to the low temperature phase, prevails even above the $T_c$(~ 150 K) with significant intensity upon heating. The complete transition takes place above ~ 230 K, as also suggested by our hysteretic Raman modes (see Fig. 6) thus further corroborating the proposition of coexistent phases across the transition. We conjecture that the nucleation for the phase transition takes place locally within the host crystal leading to the formation of micro-domains bounded by domain walls (i.e. a coexistence of both the phases) and, therefore, the complete transition occurs eventually at temperatures much above the $T_c$ ~ 150 K. This is similar to the observation of coexistence of phases in perovskite $SrTiO_3$ even above its structural phase transition at ~ 105 K[57-59]. Based on the Landau theory of phase transition, it can be proposed that a non-zero order parameter associated with the unstable phonon (soft mode Z1) with irreducible representation $\Gamma_3^+$ (involving Ba and Zn translation with $TeO_6$ octahedral rotation) in the phonon dispersion triggers the structural phase transition in BZT in concomitant with the para-to-ferroelastic transition[60].

**3.6 Phase transition and phonon band splitting: A group-subgroup transformation**

The structural transformation, from trigonal (R-3m) to monoclinic (C2/m) phase, follows Aizu's notation '-3mF2/m' in BZT. This group-to-subgroup transformation leads to a loss of point group symmetry while preserving the translational symmetry with respect to the prototypical trigonal phase[61]. An estimation of the low temperature monoclinic phase lattice parameters of BZT can be made in terms of the room temperature trigonal phase parameters using the following set of relations[62]

$$a_m = \sqrt{3}a_t, \quad b_m = a_t, \quad c_m = c_t/3 \sin\beta_m, \quad \beta = 180° - \tan^{-1}(c_t/\sqrt{3}a_t) \quad (4)$$

where, $a_m$, $b_m$, $c_m$, and $\beta$ represent the lattice parameters in the low temperature monoclinic phase while $a_t$ and $c_t$ are the high temperature trigonal lattice parameters in the hexagonal setting. As obtained from our XRD results, the lattice parameters of the trigonal phase at room temperature are $a_t = b_t = 5.8236$ Å and $c_t = 28.692$ Å. Therefore, the above set of equations yield the monoclinic lattice parameters as $a_m = 10.086$ Å, $b_m = 5.8236$ Å, $c_m = 9.0228$ Å, and $\beta = 109.36^0$. The estimation of the monoclinic lattice parameters obtained by our DFT calculations are $a_m = 10.091$ Å, $b_m = 5.8240$ Å, $c_m = 10.195$ Å, and $\beta = 109.94^0$, which are in good agreement with our experimentally derived values. The transition suggests lowering of the symmetry in the *ab*-plane as BZT transforms from trigonal to monoclinic phase. This justifies the observed splitting of most of the phonon modes with $E_g$ symmetry below the transition temperature ~ 150 K, as shown in Fig. 4. The lowering of symmetry upon transition and increase in the number of phonons can be further understood by considering the symmetry elements and Wyckoff-site splitting scheme, respectively, as described in the supplementary information.

### 3.7 Phase transition and phonons analysis: first-principles calculations

Aiming to shed light on the lattice dynamics in the $Ba_2ZnTeO_6$ system, we computed and obtained the phonon modes in the high-symmetry *R-3m* (Space group: 166) at the Γ-point (see Table I). At first, we have fully relaxed the atomic coordinates, volume, and lattice degrees of freedom. Thus, we obtained lattice parameters of $a_t = b_t = 5.827$ Å and $c_t = 28.702$ Å within the hexagonal representation which are in good agreement with our measured parameters obtained by XRD. The latter lattice parameters are equivalent to $a = b = c = 10.142$ Å and $\alpha = \beta = \gamma = 33.390°$ in its trigonal primitive cell, see Fig. 8(a). Interestingly, we observed three unstable modes, with negative frequency values by notation, associated with a 2-fold degenerated $E_g$ mode and a non-degenerate $A_{2g}$ mode. The former at $i\omega = -62.01$ cm$^{-1}$ whereas the latter at $i\omega = -24.61$ cm$^{-1}$. The

$E_g$ mode is related to an in-phase TeO$_6$ octahedral rotation with a rotation axis within the *xy*-plane (in the hexagonal reference), while the $A_{2g}$ mode corresponds to an in-phase TeO$_6$ octahedral rotation along the hexagonal *z*-axis (*i.e.* stacking axis), with Ba and Zn translations (see Table I). Our calculations suggest the $E_g$ mode as unstable driving the structural phase transition, thus in full agreement with the experimental findings discussed above. Therefore, we proceed to condense the eigen-displacements related to the $E_g$ (*i.e.* $\Gamma_3^+$ mode) into the high-symmetry structure, as shown in Fig. 8(a). This procedure is well-known to explain the source of the displacive phase transitions as in the case of ferroelectric perovskite oxide BaTiO$_3$[63], the strain-induced multiferroic state in the NaMnF$_3$ fluoride[64, 65], and the polar response in the Sr(Nb,Ta)O$_2$N[66, 67] oxynitrides. Figure 8(b) presents the total energy-wells, in meV/f.u., computed for fractional displacements of the atomic species taken along the eigenvectors' directions. These energy curves are obtained by comparing the total energy per formula-unit of the high-symmetry *R-3m* with the low-symmetry phase, calculated by introducing the atomic distortions into the cell. As it can be observed, a 4$^{th}$-order energy-well is obtained when condensing the $E_g$ mode indicating an energy gain once such a distortion is introduced into the structure. As soon as the bottom of the energy well is reached, we have confirmed that the induced low-symmetry phase belongs to the *C2/m* (Space group: 12), as shown in Fig. 8(a). The latter phase explains our experimental results that show the transition, driven by the $E_g$ - phonon mode, below T ~ 150 K.

Regarding the $A_{2g}$ mode, we have followed the same procedure for the mode's condensation, as shown in Fig. 8(b). Interestingly, we observed that such a mode presents a quasi-2$^{nd}$-order behavior at the bottom of the energy-well possibly due to its anharmonic behavior. The frequency of the $A_{2g}$ mode is hardened as soon as the $E_g$ mode is introduced into the structure at the *C2/m* phase suggesting a rather strong phonon-phonon coupling between these octahedral rotational modes, thus corroborating the observation of strong anharmonicity in our experiments. After full atomic and cell relaxation, the *C2/m* phase shows lattice parameters of $a_t = b_t = 5.826$ Å and $c_t = 28.755$ Å within the hexagonal representation. The latter lattice parameters are equivalent to $a = b = c = 10.19562$ Å and $\alpha = \beta = \gamma = 33.382°$ in the trigonal primitive cell. Table I presents the Raman phonon modes measured experimentally in the low-temperature regime, as well as computed for the low-symmetry *C2/m* phase. From the theoretical side, we observed that all of the vibrational modes are fully stable (*i.e.* positive frequency values) confirming the thermal and vibrational stability of such a phase in the $Ba_2ZnTeO_6$ system.

## 4. SUMMARY AND CONCLUSION

To summarize, we have investigated the lattice dynamics of trigonal $Ba_2ZnTeO_6$ across its ferroelastic phase transition at ~ 150 K with the help of temperature-dependent Raman spectroscopy, XRD, and specific heat measurements as well as DFT calculations. While our extensive Raman study reveals various important observations related to the phase transition such as the presence of the central peak, hysteretic phonon behavior, coexistence of both high- and low-temperature phases in the hysteretic region, and phonon anomalies arising due to very strong anharmonicity, the specific heat and temperature-dependent XRD measurements provide further evidence for the observed phase transition. We have estimated the anharmonicity of the soft mode (Z1 mode at ~ 31 cm$^{-1}$) which has been associated with the lattice instability arising from $TeO_6$ octahedral rotation along with Ba and Zn translation, as suggested and corroborated by our DFT calculations. The phase transition and the thermal responses of the soft mode are accompanied by a rise in the intensity of the central peak (quasi-elastic Rayleigh profile) near $T_c$ with a decrease in its linewidth, suggesting the presence of additional relaxation processes associated with the lattice and its dynamics. In addition, we have shown that phonons display a hysteretic behavior with temperature which not only indicates the first-order nature of the phase transition but also the presence of coexistent phases above $T_c$, as corroborated by our x-ray diffraction and specific heat measurements. Thus, we believe that our findings in $Ba_2ZnTeO_6$ will motivate further investigations of similar unexplored hexagonal systems for gaining fundamental insights into the phase transition and exploring their potential applications.

## ACKNOWLEDGEMENTS

SS acknowledges Science and Engineering Research Board (SERB) for funding through ECR/2016/001376 and CRG/2019/002668. Funding from DST-FIST (Project No. SR/FST/PSI-195/2014(C)) and Nano-mission (Project No. SR/NM/NS-84/2016(C)) are also acknowledged. SM acknowledges DRDO, India, via ACRHEM (DRDO/18/1801/2016/01038: ACRHEM-PHASE-III) for financial assistance. G.V acknowledges Institute of Eminence, University of Hyderabad (UoH-IoE-RC3-21-046) for funding and the CMSD, University of Hyderabad, for providing the computational facilities. Authors acknowledge Central Instrumentation Facility at IISER Bhopal for PXRD and PPMS research facilities. Authors thank Chandan Patra, Department of Physics at IISER Bhopal, for helping in the acquisition of the specific heat data. Calculations presented in this paper were carried out using the GridUIS-2 experimental testbed, being developed under the Universidad Industrial de Santander (SC3-UIS) High Performance and Scientific Computing Centre, development action with support from UIS Vicerrectoría de Investigación-y Extensión (VIE-UIS) and several UIS research groups as well as other funding resources. A.C.G.C. acknowledge the grant No. 2677 entitled "Quiralidad y Ordenamiento Magnético en Sistemas Cristalinos: Estudio Teórico desde Primeros Principios" supported by the VIE – UIS.

**Table I.** *Phonon frequencies (ω) and symmetry assignments of $Ba_2ZnTeO_6$ in both trigonal and monoclinic phase. Atoms involved in the phonons of the trigonal phase are listed.*

| MODE | Trigonal (R-3m) | | | | Atoms (Th.) Trigonal | Monoclinic (C2/m) | | |
|---|---|---|---|---|---|---|---|---|
| | ω (cm$^{-1}$) | | Symmetry assignment | | | ω (cm$^{-1}$) | | Symmetry Assignment (Th.) |
| | Exp. | Th. | Our work (Exp.+Th.) | Earlier report[Ref[23]] | | Exp. | Th. | |
| Z1 | 31 | -62 | $E_g$ | $E_g$ | Ba tr. + Zn tr. + TeO$_6$ rot. | 38 | 55 | $B_g$ |
| | | | | | | 50 | 75 | $A_g$ |
| Z1' | - | -25 | $A_{2g}$ | - | TeO$_6$ rot. + ZnO$_6$ rot. | - | 79 | $B_g$ |
| Z2 | 87 | 79 | $A_{1g}$ | $A_{1g}$ | Ba tr.+ Zn tr. + O tr. | 90 | 92 | $A_g$ |
| Z3 | 103 | 92 | $E_g$ | $A_{1g}$ | Ba tr.+ Zn tr. + ZnO$_6$ rot. | 99 | 97 | $A_g$ |
| | | | | | | 107 | 102 | $B_g$ |
| Z4 | 110 | 104 | $A_{1g}$ | $A_{1g}$ | Ba tr.+ Zn tr. + O tr. | 116 | 119 | $A_g$ |
| Z5 | 120 | 112 | $E_g$ | $E_g$ | Ba tr.+ Zn tr. + TeO$_6$ rot. | 119 | 121 | $B_g$ |
| | | | | | | 123 | 128 | $A_g$ |
| Z6 | 153 | 135 | $E_g$ | $A_{1g}$ | Ba tr. + TeO$_6$ rot. | 142 | 147 | $B_g$ |
| | | | | | | 163 | 160 | $A_g$ |
| Z6' | silent | 145 | $A_{2g}$ | $A_{2g}$ | TeO$_6$ rot. + ZnO$_6$ rot. | 185 | 181 | $B_g$ |
| Z7 | 382 | 340 | $E_g$ | $E_g$ | O - O sciss. | 378 | 371 | $A_g$ |
| | | | | | | 381 | 374 | $B_g$ |
| Z8 | 394 | 374 | $E_g$ | $E_g$ | O - O sciss. | 396 | - | - |
| | | | | | | 397 | - | - |
| Z9 | 405 | - | $A_{1g}$ | $E_g$ | - | 405 | 375 | $A_g$ |
| Z10 | 470 | 436 | $A_{1g}$ | $A_{1g}$ | Zn trans. + O-O sym. str. | 469 | 436 | $A_g$ |
| Z11 | 573 | 559 | $E_g$ | $E_g$ | O-O sym. str. + O-O asym. str. | 573 | 558 | $B_g$ |
| | | | | | | 576 | 561 | $A_g$ |
| Z12 | 616 | 603 | $E_g$ | $E_g$ | O-O sym. str. + O-O asym. str. | 616 | 587 | $A_g$ |
| | | | | | | 621 | 588 | $B_g$ |
| Z13 | 689 | 670 | $A_{1g}$ | $E_g$ | O-O sym. str. | 692 | 673 | $A_g$ |
| Z14 | 736 | 726 | $A_{1g}$ | $A_{1g}$ | O-O sym. str. | 735 | 705 | $A_g$ |
| Z15 | 766 | - | $E_g$ | $E_g$ | - | 767 | - | - |
| | | | | | | 770 | - | - |

**Exp. – Experiment, Th. – Theory, tr. – translation, rot. – rotation, sciss. – scissoring, sym. str. - symmetric stretching, and asym. str. - asymmetric stretching**

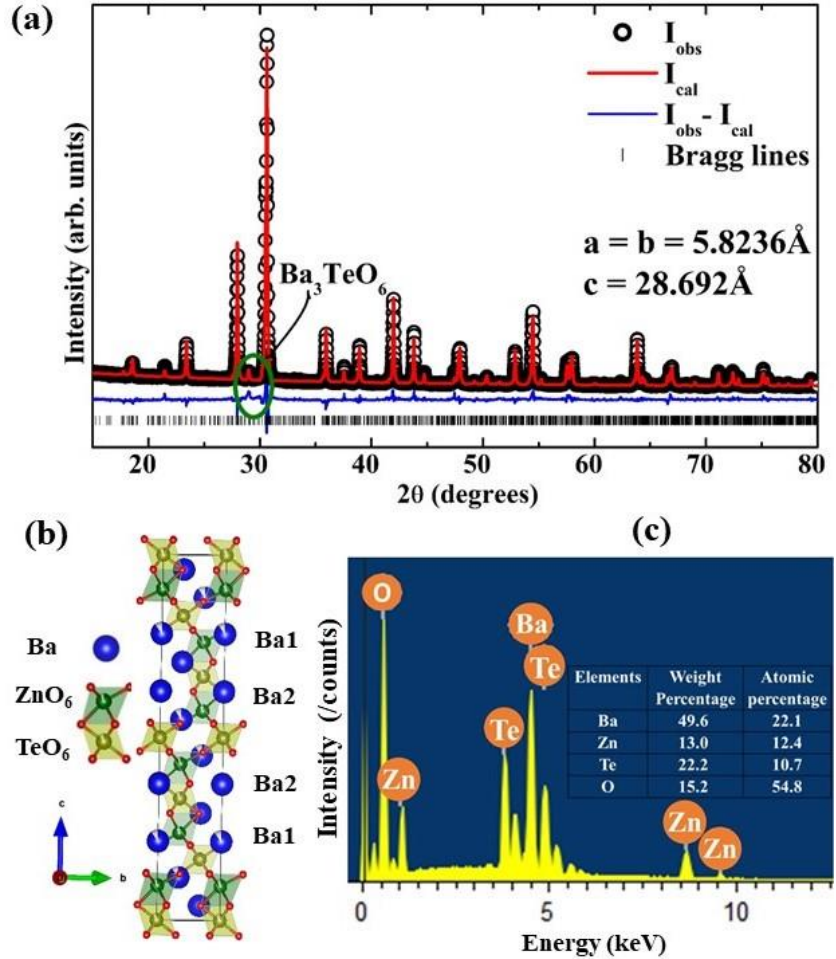

**Fig. 1** (Color online)**: (a)** Room-temperature x-ray diffraction pattern of $Ba_2ZnTeO_6$ showing the lattice parameters derived from the refined diffraction profile where $I_{obs}$ is the observed experimental data (black open circles), $I_{cal}$ is the fitted (calculated) x-ray profile (red solid line), and their difference, i.e. $I_{obs} - I_{cal}$ is represented with the blue solid line. The reflection peak due to the presence of a secondary phase of $Ba_3TeO_6$ (of ~ 2.1 %) is encircled with a green circle. **(b)** Trigonal crystal structure of BZT showing the face shared $ZnO_6$-$TeO_6$ octahedra. **(c)** EDAX analysis showing the atomic and weight percentage of the elements present in BZT.

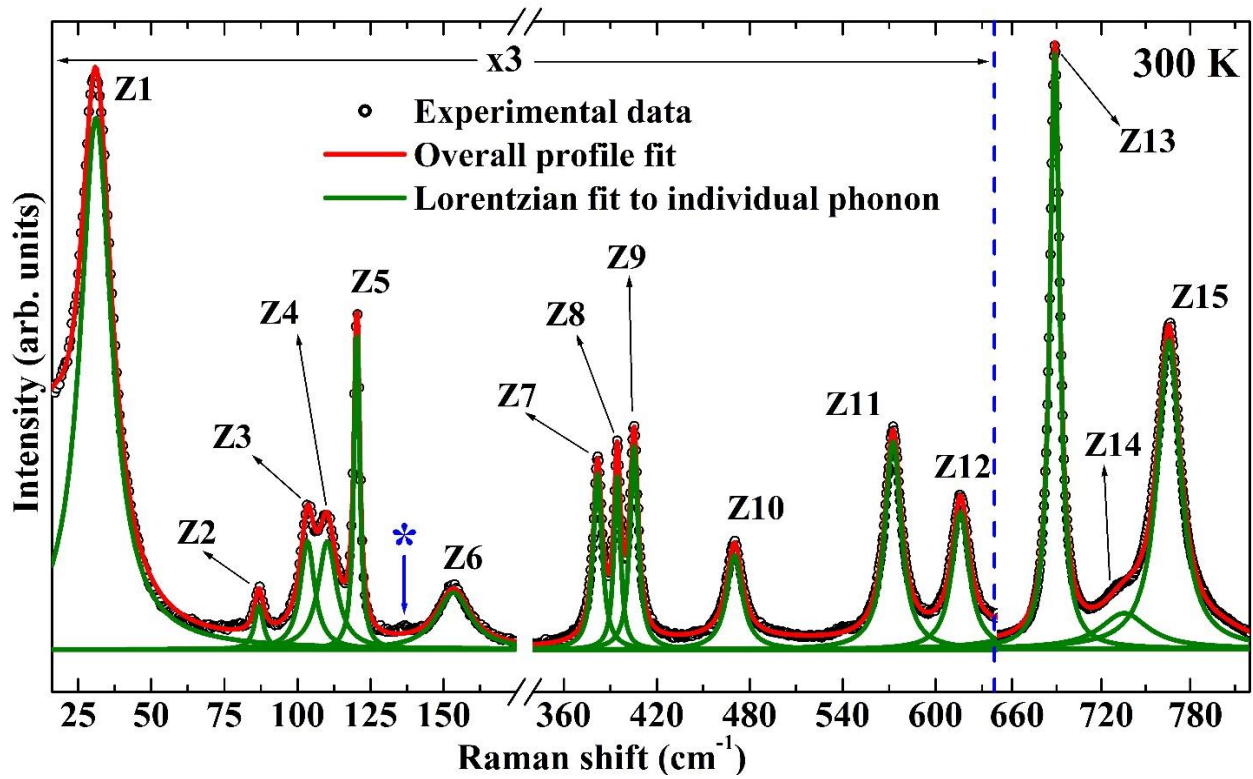

**Fig. 2** (Color online)**:** Room-temperature Raman spectrum of $Ba_2ZnTeO_6$ with modes labelled as Z1 to Z15. Modes below ~ 650 $cm^{-1}$ are scaled for clarity. The weak peak at 135 $cm^{-1}$ labelled with asterisk (*) is probably the missing $A_{1g}$ mode, as discussed in the main text.

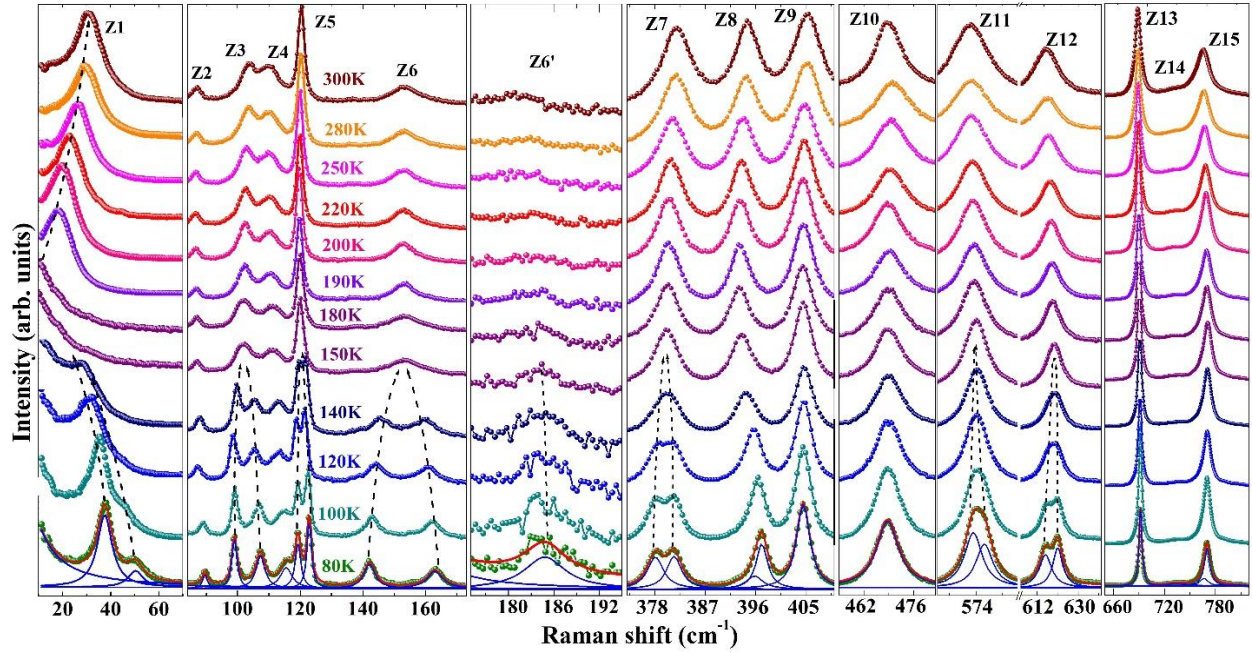

**Fig. 3** (Color online)**:** Raman spectra of $Ba_2ZnTeO_6$ shown at a few temperatures across the phase transition (at $T_c \sim 150$ K). Dashed arrows are guide to the eye representing phonon splitting and shift of the modes.

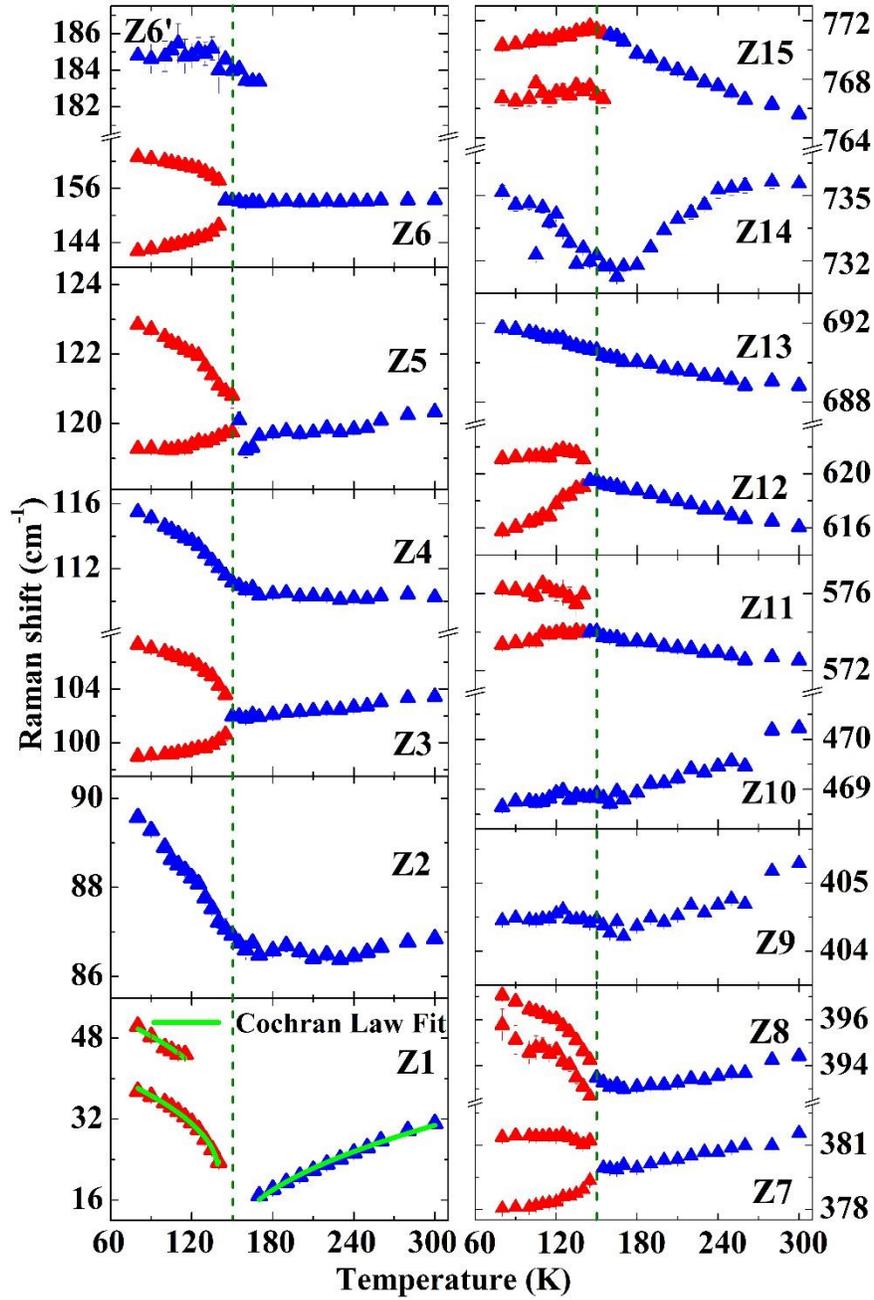

**Fig. 4** (Color online)**:** Evolution of the phonon frequencies as a function of temperature exhibiting splitting of the modes upon cooling below the transition as well as signatures of strong anharmonicity, as discussed in the text. Vertical dashed lines represent the transition temperature ($T_c \sim 150$ K). The solid lines for the mode Z1 show the fitting by Cochran relation (Eqn. 2).

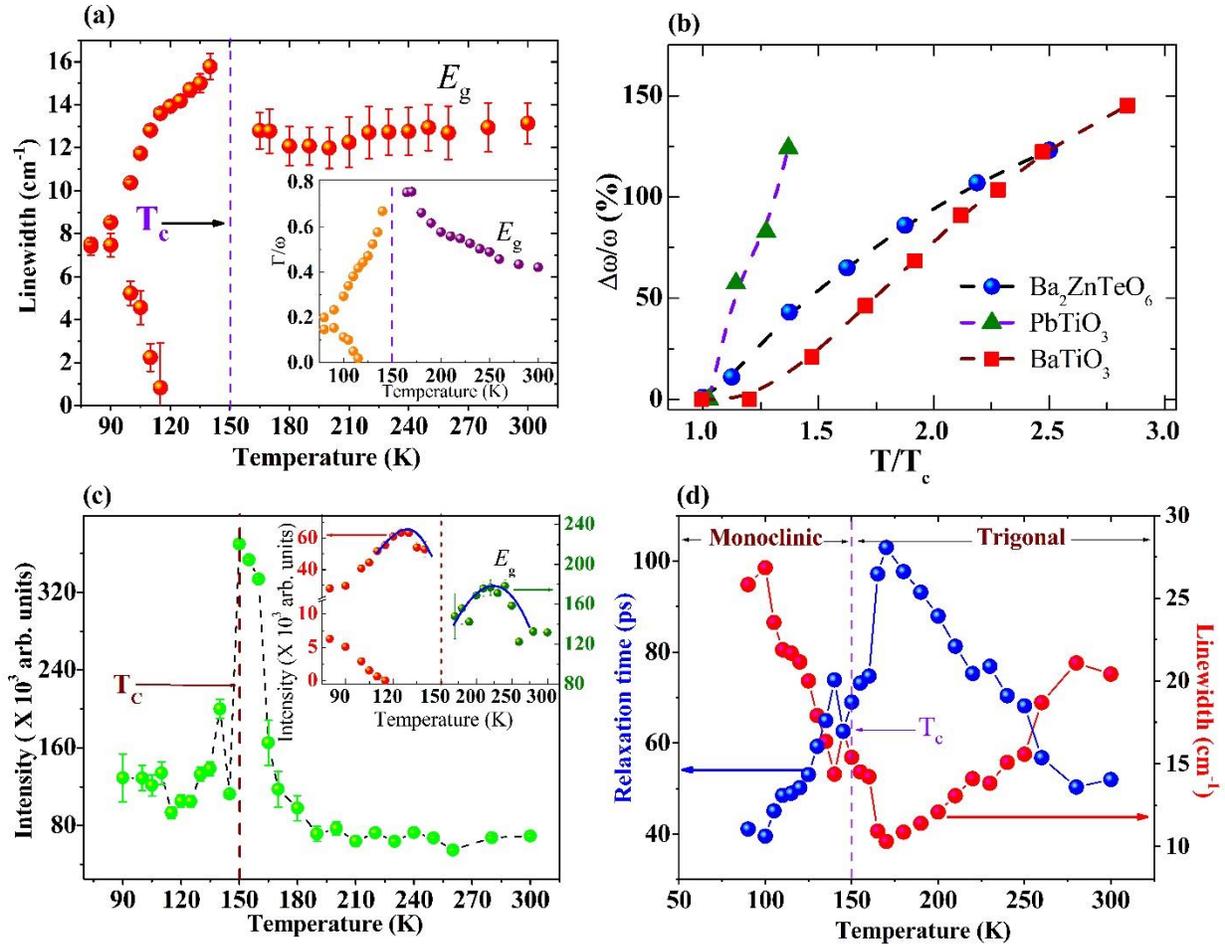

**Fig. 5** (Color online)**: (a)** Linewidth of the soft-mode as a function of temperature. Inset shows the damping $\left(\frac{\Gamma}{\omega}\right)$ of the soft mode across the ferroelastic transition, **(b)** Comparison of the soft mode (Z1) frequency-shift $\left(\frac{\Delta\omega}{\omega}\,\%\right)$ of $Ba_2ZnTeO_6$ with classic perovskite systems $BaTiO_3$ [45] and $PbTiO_3$ [46] as a function of temperature normalized with respect to the respective $T_c$, **(c)** Variation of the intensity of the central peak as a function of temperature where the vertical dashed line represents the transition temperature. The inset shows the variation of the intensity of the soft mode (Z1) with temperature where solid lines (in blue color) represent guide to eye for the decreasing intensity near $T_c$, and **(d)** Variation of relaxation time and linewidth of the central peak over temperature where dotted line represents the transition temperature.

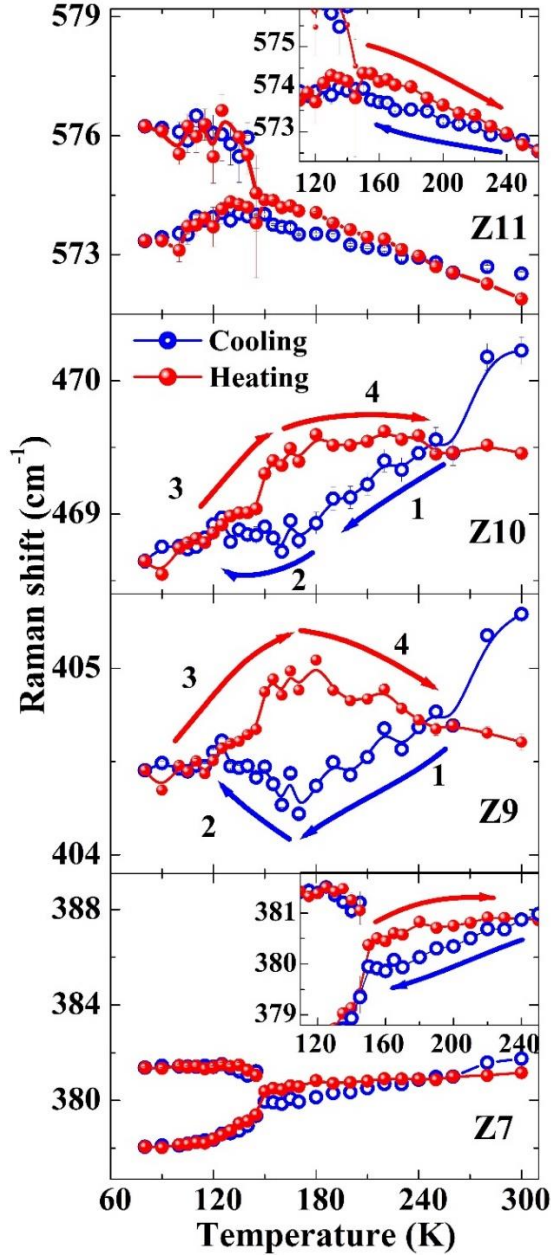

**Fig. 6** (Color online)**:** Hysteretic shift in the phonon frequencies of Z7, Z9, Z10, and Z11 during cooling (blue open circles) and heating (red filled circles) cycles. Insets are shown for visual clarity of the hysteresis. Error bars represent the standard deviation in the phonon frequencies which are smaller than or comparable to the symbol size.

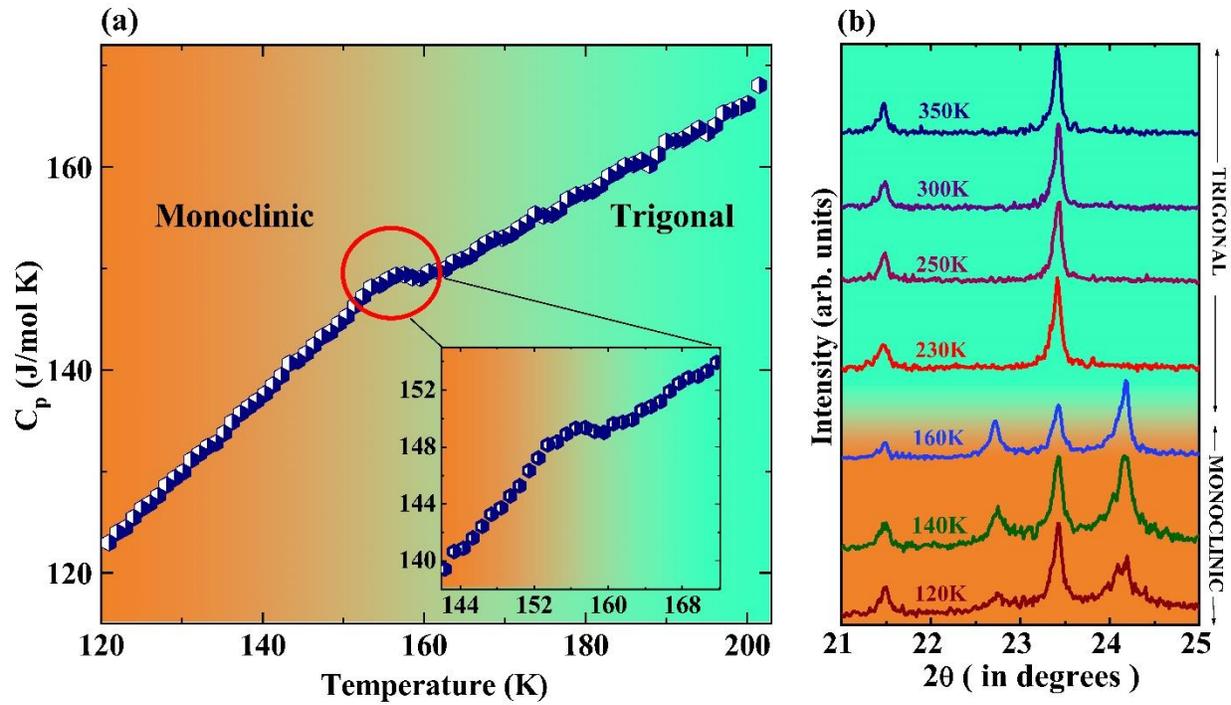

**Fig. 7** (Color online)**: (a)** Specific heat of $Ba_2ZnTeO_6$ displaying a weak anomaly (magnified in the inset) near the phase transition **(b)** X-Ray diffraction profiles of $Ba_2ZnTeO_6$ at a few temperatures showing emergence of new reflections in the low-temperature monoclinic phase.

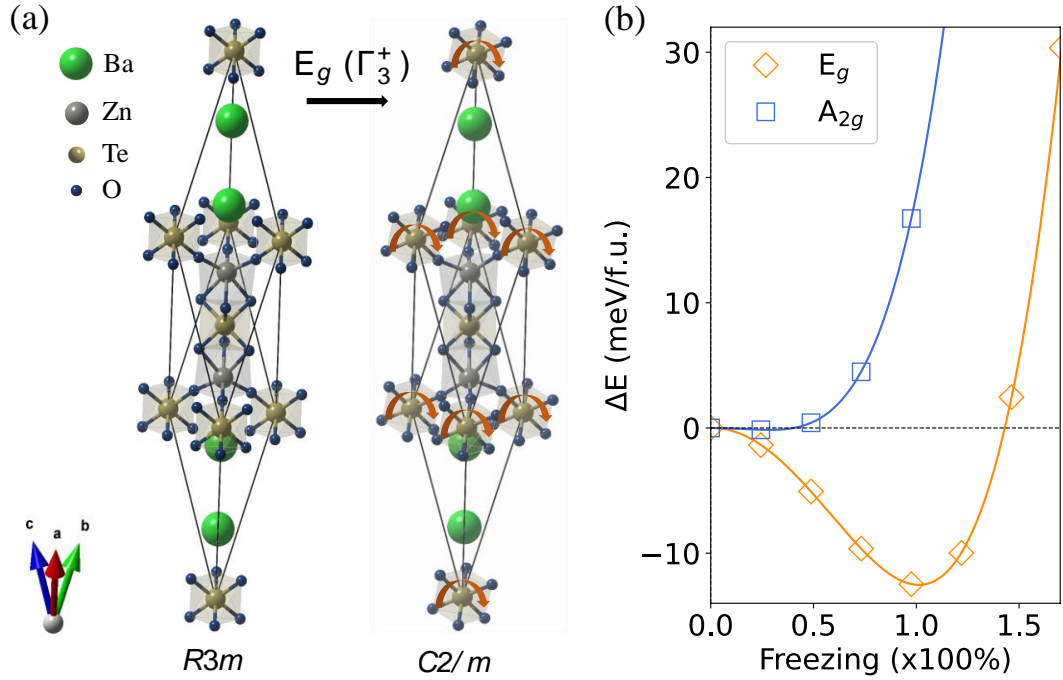

**Fig. 8** (Color online): **(a)** High-symmetry rhombohedral structure, *R-3m* (SG. 166) as well as the low-symmetry *C2/m* (SG. 12) phase obtained after the $E_g$ ($\Gamma_3^+$) phonon condensation. As it can be observed, the phonon mode is associated with a TeO$_6$ octahedral rotation, marked by the orange arrows, in the Ba$_2$ZnTeO$_6$ compound. **(b)** Total energy-well profiles, in meV/f.u. computed for the $E_g$ and $A_{2g}$ modes condensation into the *R-3m* high-symmetry structure.

# Supplementary Information

# Lattice dynamics across the ferroelastic phase transition in $Ba_2ZnTeO_6$:

# A Raman and first-principles study

Shalini Badola[1], Supratik Mukherjee[2], B. Ghosh[1], Greeshma Sunil[1], G. Vaitheeswaran[3], A. C. Garcia-Castro[4], and Surajit Saha[1]*

[1]*Indian Institute of Science Education and Research Bhopal, Bhopal 462066, India*

[2]*Advanced Center of Research in High Energy Materials (ACRHEM), University of Hyderabad, Prof. C. R. Rao Road, Gachibowli, Hyderabad 500046, Telangana, India*

[3]*School of Physics, University of Hyderabad, Prof. C. R. Rao Road, Gachibowli, Hyderabad 500046, Telangana, India*

[4]*School of Physics, Universidad Industrial de Santander, Calle 09 Carrera 27, Bucaramanga, Santander, 680002, Colombia*

*\*Correspondence: surajit@iiserb.ac.in*

This supplementary information contains details of mode symmetry assignments and eigenvectors of the phonon modes, high-temperature Raman data, thermal response of lattice constants, linewidth of soft mode and central peak, as well as specific heat data and Wyckoff-site splitting scheme.

## S1. Temperature dependence of the Raman spectrum (80 – 400 K)

### (a) Mode symmetry assignments

Figure S1 depicts the evolution of the phonon bands as temperature progresses from 80 to 400 K. As it is evident from Fig. S1(a), Z1 initially demonstrates an anomalous trend (i.e. a decrease in the frequency with decreasing temperature) upon cooling and switches to a normal behavior (i.e. an increase in the frequency with lowering temperature) below 150 K. Besides, it splits into two phonon bands below the transition temperature, thus lifting its in-plane $E_g$ symmetry (double-degeneracy) to $A_g$ and $B_g$ symmetries at low temperatures. Such behavior (i.e. softening and vanishing of a mode) is a typical characteristic of soft phonon that plays an important role in driving a system to an ordered phase (ferroelectric/ferroelastic) of displacive-type[1,2]. Notably, the disappearance of the soft mode (before splitting) upon cooling down to ~ 170 K (i.e. merging with the central peak) is an indicative of a first-order phase transition. Our DFT-based calculations further suggest that the soft phonon (Z1 mode) involving the displacement of Ba and Zn atoms along with $TeO_6$ octahedral rotation in BZT gives rise to lattice instabilities leading to a transition from the high-temperature trigonal to low-temperature monoclinic phase.

Signatures of the phase transition are also evident in other phonon modes of BZT in the form of phonon splitting and unusually large shifts below $T_c$, as shown in Fig. S1(b). The modes Z2 and Z4 shift rapidly with temperatures below ~ 150 K as compared to their shifts at higher temperatures (see Fig. S1(b) and Fig. 4 in the main text), providing another indication of the phase transition. While the observed behavior for Z2 is in agreement with an earlier report by Moreira *et al.*[3], Z4 shows a disparity in its thermal behavior below the transition temperature where we observe a larger shift with temperature. The modes Z2 and Z4 can be assigned to $A_g$ symmetries. The mode Z3 splits into two components at lower temperatures and, hence, at room temperature, it can be assigned to $E_g$ symmetry. Similarly, we observe the splitting of the Z5, Z6, Z7, and Z8 phonon bands into two modes for each of them upon cooling below the transition temperature (~ 150 K). Therefore, all these modes (Z5 to Z8) at room temperature can be assigned to $E_g$ symmetries. According to Moreira *et al.*[3], the modes observed at ~142 cm$^{-1}$ and ~163 cm$^{-1}$ (in our case these are split components of Z6) at 80 K distinctly exist up to room temperature without merging which is in contrast to our observation. We have observed that these modes persist till the phase transition upon heating and merge into a single-mode that is identified as Z6 (at ~153 cm$^{-1}$) above ~ 150 K, as shown in Fig. S1(b). Therefore, we assign Z6 to $E_g$ symmetry contrary to the earlier assignment of $A_{1g}$ symmetry[3]. As shown in Fig. 3 in the main text, the mode Z6′ is clearly visible in the low-temperature monoclinic phase possessing a normal shift in frequency with temperature. However, the intensity of Z6′ becomes extremely weak to be deciphered at higher temperatures (in the trigonal phase). Therefore, based on its behavior and an earlier report[3], it can be assigned to $B_g$ symmetry. Moreira *et al.*[3] observed a bifurcation of the mode Z9 (at ~ 405 cm$^{-1}$) near the phase transition and hence, assigned it to $E_g$ symmetry. However, we note that Z9 displays a weak but anomalous temperature-dependent shift with a change in slope across the phase transition temperature without undergoing any splitting at lower temperatures. Importantly, the polarization-dependent Raman data by Moreira *et al.* demonstrate a considerable decrease in the mode intensity of Z9 in cross-polarization, which is suggestive of its $A_{1g}$ ($A_g$ below $T_c$) symmetry counter to their assignment of $E_g$ symmetry[3]. Therefore, we assign the Z9 to $A_{1g}$ symmetry. A similar disagreement was again noticed with the symmetry assignment of the phonon mode Z13 which can again be resolved by considering their polarization-dependent Raman data that show a significant decrease in the Raman intensity upon changing the polarization from cross to parallel configuration. Thus, we assign the mode Z13 to $A_{1g}$ symmetry. The mode Z10 exhibits an anomalous shift with temperature without undergoing splitting and, therefore, it is assigned to $A_{1g}$ symmetry. The modes Z11, Z12, and Z15 split below the phase transition temperature. Therefore, we assign them (Z11, Z12, and Z15) as $E_g$ symmetry modes. Notably, the mode Z14 ($A_{1g}$) shows a change in behavior (slope) at ~ 150 K without undergoing any splitting of the band, which is in agreement with the earlier report[3].

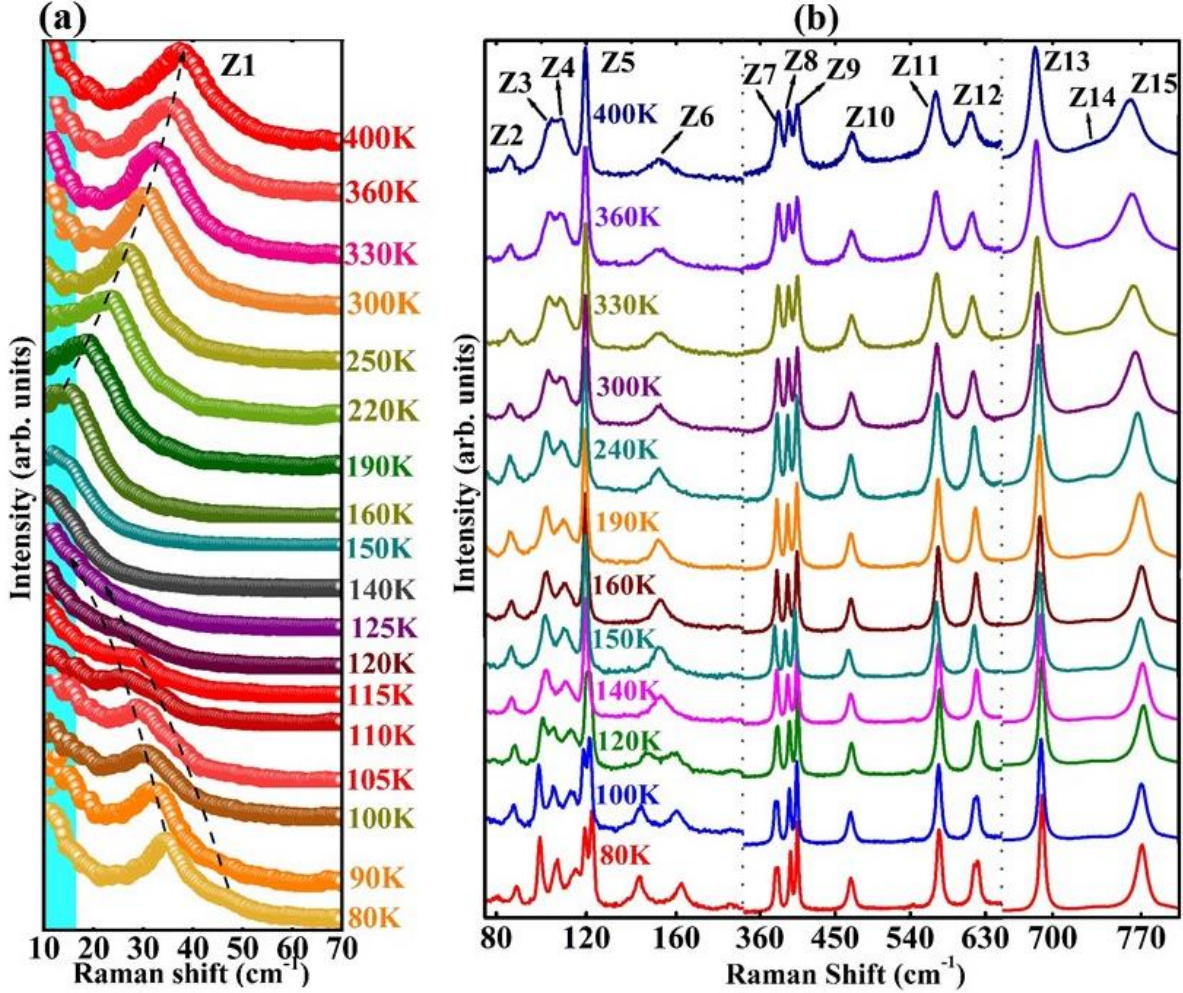

**Figure S1: (a)** *Evolution of soft phonon (Z1) as a function of temperature where dotted arrows represent a guide to the eye and the shaded region in blue indicates signatures of the central peak.* **(b)** *Raman spectra of $Ba_2ZnTeO_6$ at temperatures up to 400 K. Dotted lines separate different regions of frequency which are scaled accordingly for showing the low-intensity peaks and phonon splitting at low temperatures.*

**(b) Phonon eigenvectors**

The eigenvectors corresponding to the Raman active phonon vibrations in the trigonal phase of $Ba_2ZnTeO_6$ are shown in Fig. S2 below. Based on our DFT calculations, we find negative phonon modes Z1 and Z1′ at -62 cm$^{-1}$ ($E_g$) and -24 cm$^{-1}$ ($A_{2g}$) (Raman inactive in trigonal phase), respectively. It can be observed that the $E_g$ mode arises due to $TeO_6$ rotation along with translation of Ba and Zn atoms whereas the $A_{2g}$ mode originates from the rotation of both $ZnO_6$ and $TeO_6$ octahedra. As we move towards the positive frequencies (Z2 – Z6′), the translational motion of all atoms is seen accompanied by $TeO_6$ and $ZnO_6$ octahedral rotations as described in Table I of the main text. Further on moving towards the higher frequency modes (Z7 – Z15), we observe scissoring motion, and symmetric and asymmetric stretching of the O atoms. However, the phonon modes Z9 and Z15 could not be assigned theoretically in contrast to experiments in the trigonal phase.

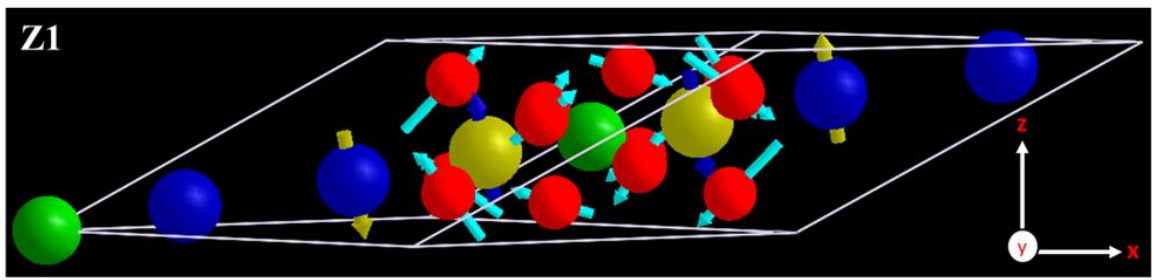

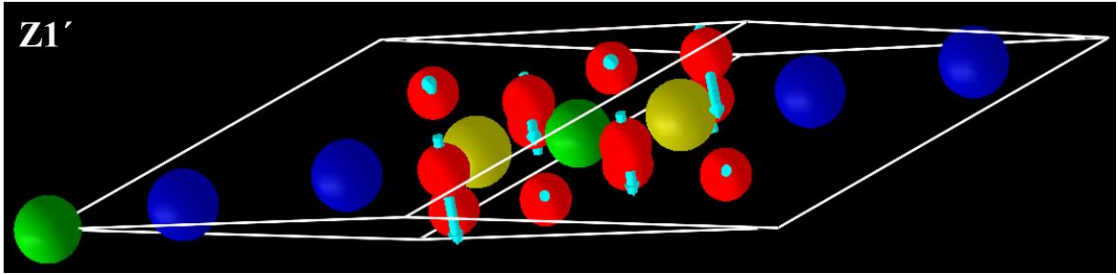
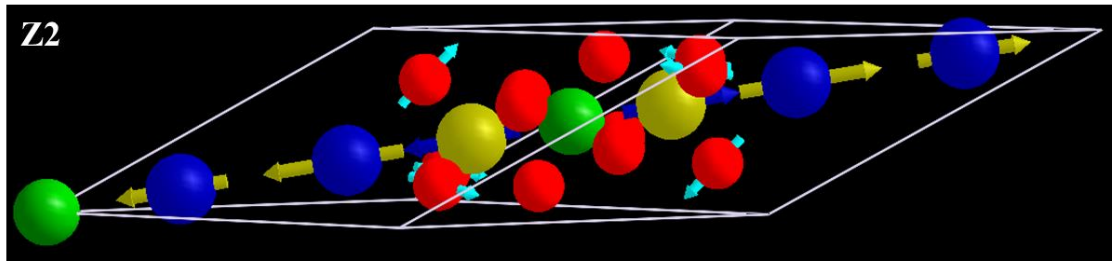
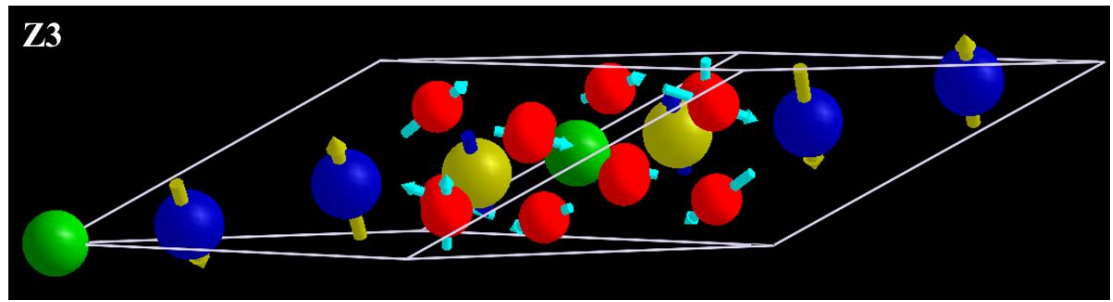
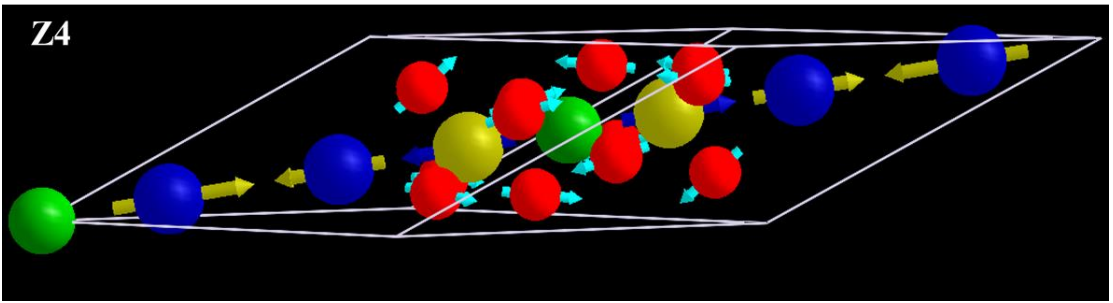
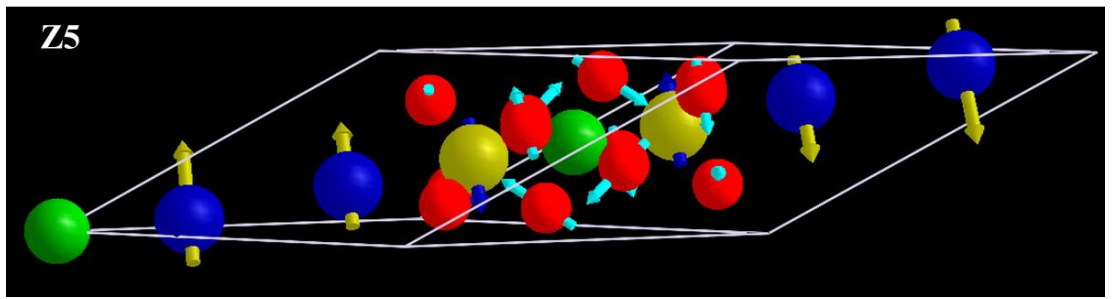

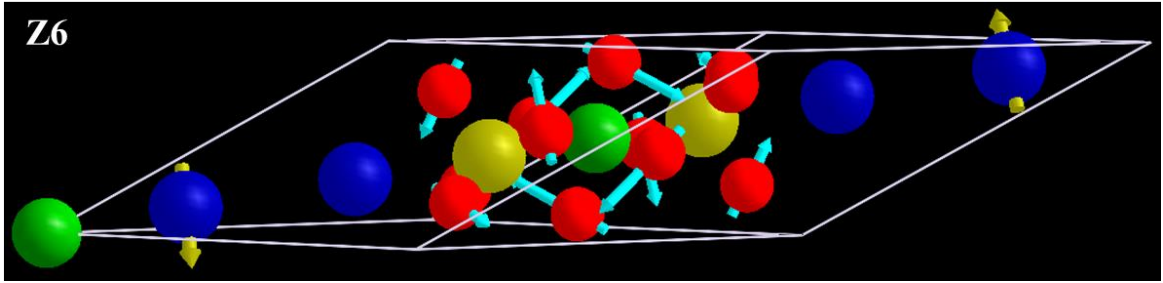
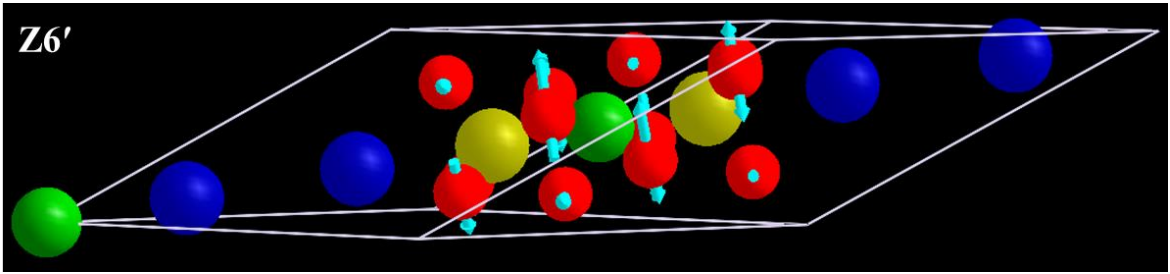
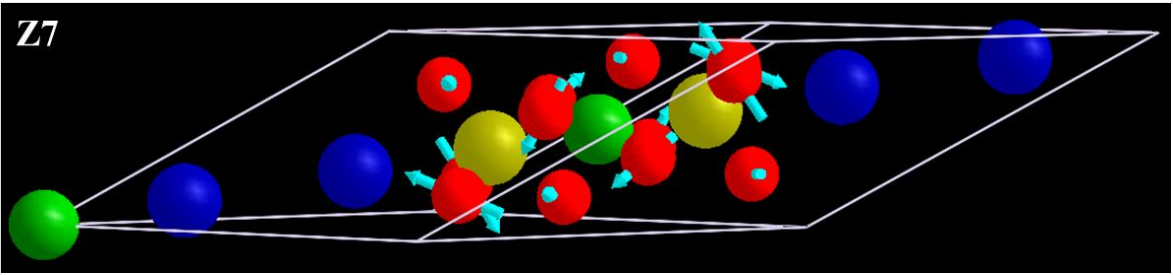
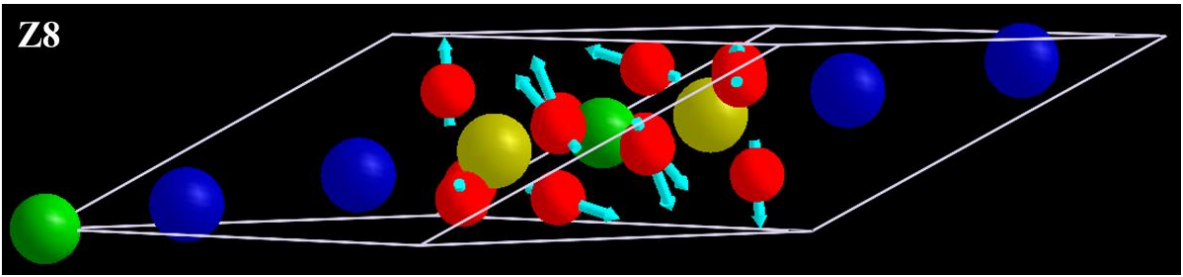
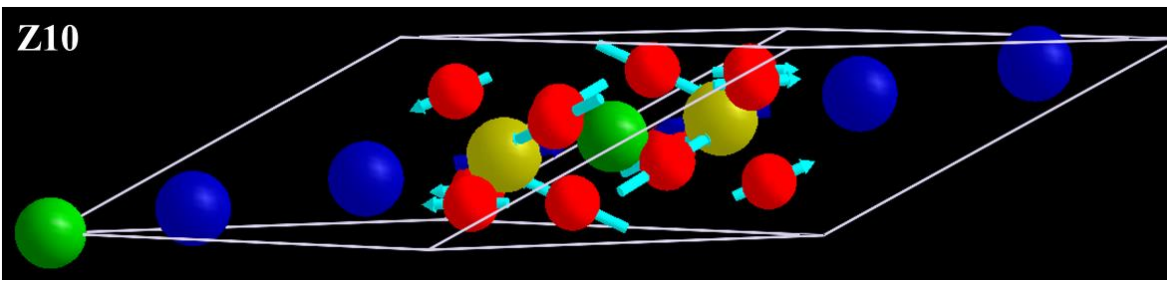

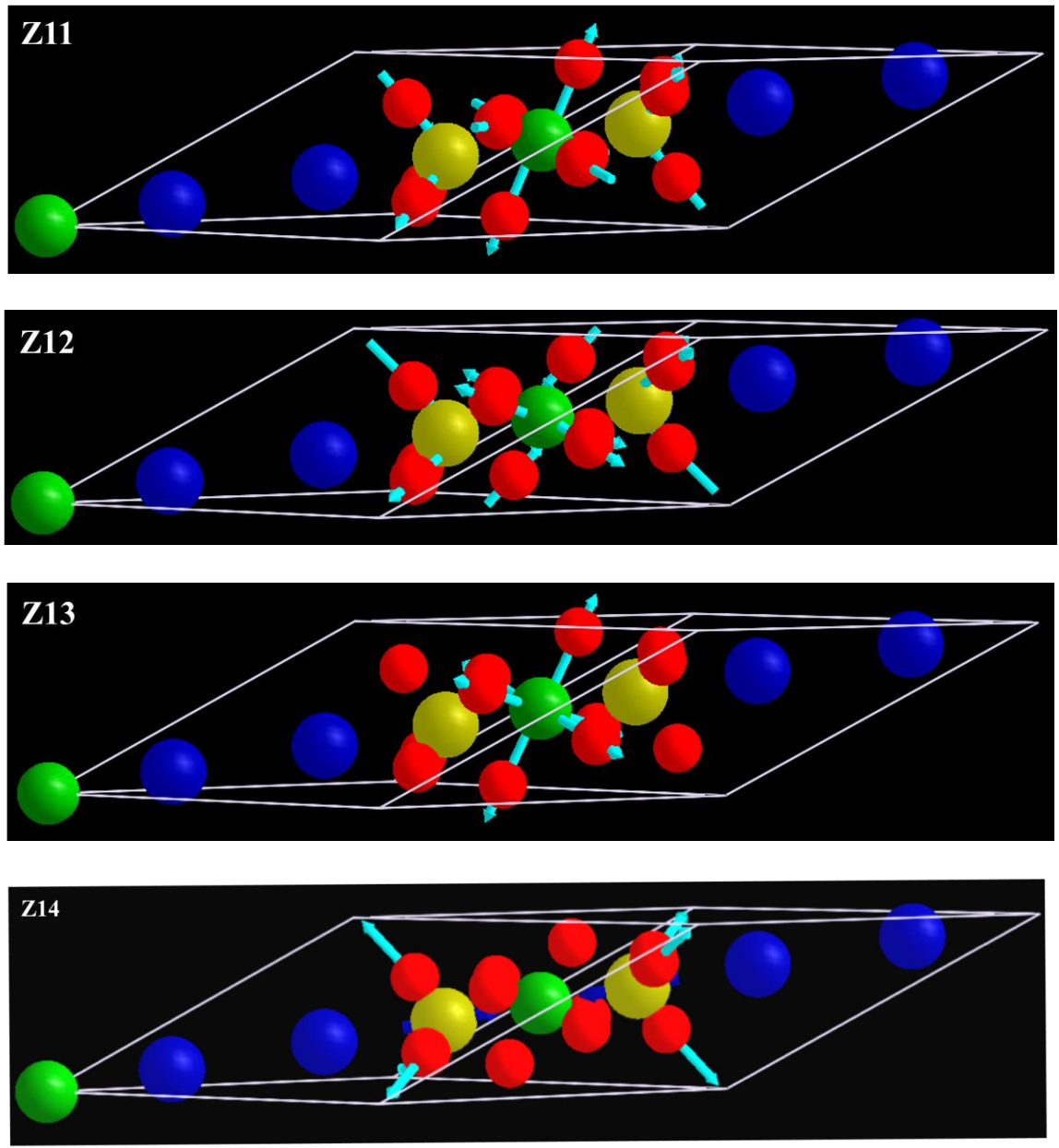

**Figure S2:** *Eigenvectors for the phonon modes of Ba$_2$ZnTeO$_6$ in the trigonal phase. The atoms shown with blue, green, yellow, and red colors are Ba, Te, Zn, and O atoms, respectively. Arrows are shown to indicate the direction of vibration of atoms.*

**(c) Quasi-harmonic and phonon-phonon anharmonic interactions**

As we notice the frequency response of the soft mode Z1 over-temperature, remarkably large shift in frequency was observed than usually expected. An unusually large shift in the frequency of the soft mode over temperature can be attributed to the dominant phonon-phonon anharmonic interactions. The quasi-harmonic change in frequency can be obtained by estimating the change in lattice volume $\left(\frac{\Delta V}{V}\right)$ over temperature, as given below. As discussed in the main text, ruling out the contributions from spin-phonon and electron-phonon interactions in non-magnetic and insulating BZT, the overall shift in phonon frequency ($\Delta\omega(T)$) as a function of temperature can be attributed to quasi-harmonic (volume-dependent) and intrinsic anharmonic (phonon-phonon) interactions.

On examination of tellurium-based systems like $Sr_2ZnTeO_6$ [4], the Grüneisen parameters ($\gamma$) for different modes were found to be typically in the range of ~ 0.9 – 2, suggesting a small contribution to the frequency shift from the quasi-harmonic term. Therefore, we expect the quasi-harmonic contribution in BZT to be comparable to that in $Sr_2ZnTeO_6$. An estimate of the frequency shift ($\Delta\omega^i \sim \Delta\omega_{qh}(T)$) produced by volume-dependent quasi-harmonic term for a phonon mode '$i$' can be made using the Grüneisen relation

$$\frac{\Delta\omega^i}{\omega^i} = -\gamma^i \frac{\Delta V}{V} \qquad [1]$$

where, $\gamma^i$ is the mode Grüneisen parameter, $\frac{\Delta V}{V}$ represents the relative change in volume which is extracted from XRD data and $\frac{\Delta\omega^i}{\omega^i}$ represents the relative change in frequency when the temperature is varied.

In the pure trigonal phase (above 230 K), the relative change in volume $\left(\frac{\Delta V}{V}\right)$ of the unit cell over temperature is ~ 0.5 % (discussed below). Assuming $\gamma^i \sim 1$ in Eq. 1, the quasi-harmonic contribution $\Delta\omega_{qh}(T)$ is evaluated to be roughly around 0.14 cm$^{-1}$ $\left(\frac{\Delta\omega_{qh}}{\omega} \sim 0.5\ \%\right)$ for the high-temperature branch of the soft mode (Z1). Notably, the quasi-harmonic contribution is negligibly small as compared to the overall shift in the frequency in the pure trigonal phase. This implies that the volume-dependent part has a negligible contribution to the observed phonon shift. As a result, it can be proposed that the unusually large shift in the phonon frequency originates entirely due to strong phonon-phonon anharmonic interactions. Thus, in the present scenario, the anharmonicity can be estimated simply as the relative change in frequency $\frac{\Delta\omega}{\omega} = \frac{\omega(T_{max}) - \omega(T_{low})}{\omega(T_{low})}$, where $T_{max}$ is the maximum temperature recorded (400 K in our case) and the lowest frequency ($\omega(T_{low})$) of the soft mode is taken at 170 K (in the high-temperature phase) since the soft mode is not resolvable at $T_c \sim 150K$. An estimation of the anharmonicity $\left(\frac{\Delta\omega}{\omega}\right)$ using the temperature-dependent shift in frequency yields a value of ~ 120 % (between 170 and 400 K) for the Z1 mode which is very large as compared to the typically observed quasi-harmonic shifts of ~ 1-2 %.

Similarly, the anharmonicity of its low-temperature split components is ~ 145 % for $B_g$ and ~ 94 % for $A_g$ mode below $T_c$ down to 80 K. This suggests that the soft mode (Z1) exhibits a very strong anharmonicity throughout the recorded temperature range. The origin of this strong anharmonicity can be associated to the lattice instability near $T_c$ rendered by $TeO_6$ octahedral rotations and its coupling to the Z1′ phonon mode as suggested by DFT calculations.

## S2. Variation of lattice parameter with temperature above the transition

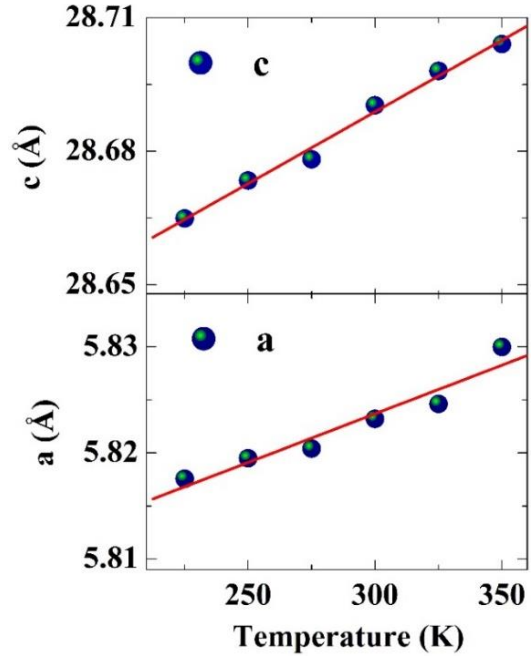

**Figure S3:** *Variation of lattice parameters of $Ba_2ZnTeO_6$ as a function of temperature where solid lines represent the linear fits to the experimental data.*

The Rietveld refinement of the x-ray diffraction profile at room temperature confirms a trigonal structure of the unit cell with lattice constant values as $a = 5.8236$ Å and $c = 28.692$ Å as shown in Fig. 1 of the main text. Figure S3 presents the variation of lattice constants as the temperature is increased in the region where the crystal is present in the pure trigonal phase. From Fig. S3, it is clear that both the lattice constants ($a$ and $c$) increase with increasing temperature indicating a positive thermal expansion of the lattice above $T_c$. It can be observed that the lattice parameters, $a$ and $c$, exhibit a change of ~ 0.2 % and ~ 0.1 %, respectively, with the varying temperature that corresponds to a change in volume by ~ 0.5 %. This, therefore, would correspond to a negligibly small quasi-harmonic contribution (0.5 %) to the phonon shifts thus justifying our attribution of the unprecedented phonon shift (in Z1 mode) to a very strong phonon anharmonicity as discussed above.

## S3. Signatures of the central peak

In addition to the observed strong anharmonicity of the soft mode, the low-frequency region near the Rayleigh scattered profile (central peak) of BZT shows an anomalous rise in intensity around the transition temperature ($T_c$ ~ 150 K). The active lattice dynamical processes at lower frequencies across the $T_c$, especially in ferroic systems exhibiting soft mode, may lead to an anomalous rise in the intensity of the Rayleigh profile i.e. the spectral region close to the excitation wavelength (near ~ 0 cm$^{-1}$), a characteristic feature which is known as central peak[5]. A comparative analysis of the frequency ($\omega$) and linewidth ($\Gamma$) (shown in the inset of Fig. 5(a) in main text) of the soft mode suggests that the mode is underdamped (because $\frac{\Gamma}{\omega} < 1$) away from $T_c$ though the damping $\left(\frac{\Gamma}{\omega}\right)$ of both the $A_g$ and $E_g$ modes increases on approaching the transition temperature. From the Raman spectra recorded at various temperatures near the structural transition, it appears that the intensity of the soft mode is transferred (spectral weight transfer) to the central peak. Thus, the soft mode is completely masked by the central peak and we observe a sharp rise in the intensity (spectral weight) of the central peak which is evident from Fig. S1(a) where the central peak is marked with shaded background in the lowest-frequency region.

Upon careful examination, we observe that the relaxation time (linewidth) of the central peak increases (decreases) near $T_c$ as shown in Fig. 5(d) in main text. A simultaneous increase in the intensity of the central peak near $T_c$ (see Fig. S1(a)) indicates the involvement of relaxation mechanisms[6-9]. The relaxation time ($\tau$) is derived from the relation $\tau = \frac{1}{\pi c \Gamma}$ where 'c' is the speed of the light and $\Gamma$ represents the linewidth of the central peak[10]. It is to be noted that the behavior of linewidth and relaxation time is derived for a qualitative understanding of the mechanisms involved and not for a quantitative analysis due to an instrumental limitation on the lowest measurable frequency (~ 8 cm$^{-1}$) and associated instrumental broadening of the spectral linewidth. An increase in the relaxation time indicates slowing down of the lattice dynamical processes near the phase transition owing to the dynamical lattice defects, phonon density, and entropy fluctuations as well as the motion of the domain walls in the coexisting phases (in the hysteretic region), as also discussed in the main text.

## S4. Specific heat measurement

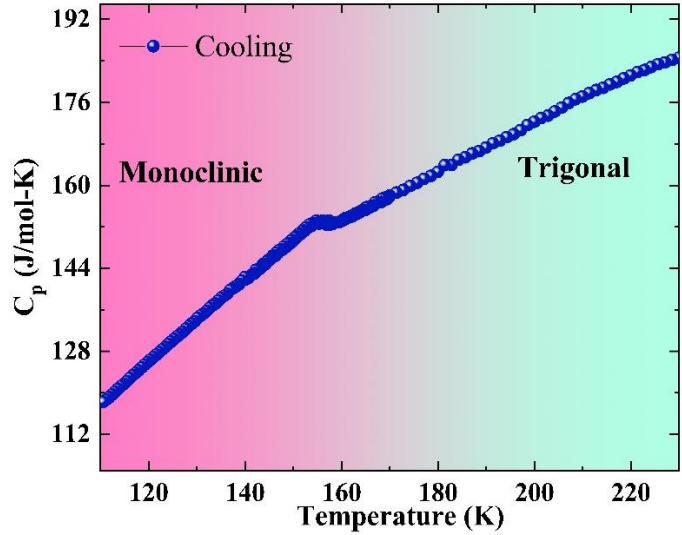

**Figure S4:** *Specific heat as a function of temperature showing a broad anomaly at ~ 150 K.*

In order to avoid any ambiguity regarding the smeared feature observed in the specific heat data, the measurement was carried out in the range of 110 – 230 K (as shown in Fig. S4) in the cooling cycle with a step size of ~ 0.25 K in between 140 and 170 K wherein a sharp transition has been observed in the Raman measurement through the splitting of some of the phonon bands. Notably, other phonon modes exhibit hysteretic thermal response, suggesting the presence of coexisting phases. We observed that the smeared transition (in specific heat) is consistent even with smaller step size confirming our conjecture that both the high (trigonal) and low temperature (monoclinic) phases coexist with each other averaging out the presence of an anticipated sharp feature in the specific heat in this temperature range.

## S5. Group-subgroup symmetry transformation and Wyckoff-site splitting scheme

BZT undergoes a structural transition from the high-symmetry trigonal to the low-temperature monoclinic phase. The high-temperature trigonal phase comprises of a total of 12 symmetry elements (E, $2C_3$, $3C_2$, i, $2S_6$, $3\sigma_d$) that define the point group '-3m' while the low-temperature monoclinic phase contains only 4 symmetry components (E, $C_2$, i, $\sigma_h$). A descent in the symmetry can be clearly marked for the low-temperature phase with a decrease in the number of symmetry components. Further, an increase in the number of phonons in the monoclinic phase can be described with the Wyckoff-site splitting scheme for the group-subgroup (-3m → 2/m) chain. In the trigonal (high temperature) phase of BZT, atoms occupy the 3a, 3b, 6c, and 18h crystallographic Wyckoff sites. Upon transition to the monoclinic (low temperature) phase, these sites are observed to transform according to the scheme presented in Table SI. It can be noted that the high-temperature Wyckoff positions 3a and 3b transform into 2a and 2c, respectively, in the low-temperature phase. According to group-theoretical analysis, 3a(2a), and 3b(2c)-sites occupied by non-magnetic Te atoms do not give rise to Raman active phonons in either the trigonal or monoclinic phases[11]. In contrast, the Ba(1), Ba(2), and Zn atoms, located at the 6c-site, lead to six Raman active phonons in the high symmetry (trigonal) phase. The 6c position transforms into the 4i-site in the low-symmetry monoclinic phase, giving rise to a total of nine modes. Further, the 18h-site of the high-symmetry trigonal phase, which leads to ten Raman active phonons, is occupied by the oxygen atoms O(1) and O(2). The 18h crystallographic position splits into 4i and 8j sites giving rise to a total of 18 phonons in the low-symmetry monoclinic phase. Thus, the low-temperature monoclinic phase predicts a total of 27 phonons with $A_g$ and $B_g$ symmetries as compared to 16 phonon modes with $A_{1g}$ and $E_g$ symmetries in the high-temperature trigonal phase. As can be noted in our Raman data, we have clearly observed 15 modes (and one very weak $A_{1g}$ mode owing to low scattering cross-section) in the high-temperature phase while 25 modes in the low-temperature phase. The two missing phonon modes in Raman data in the low-temperature phase are likely to be of $A_g$ and $B_g$ symmetries and are not vividly present possibly due to their low scattering cross-section.

**Table SI.** *Wyckoff-site splitting scheme for a phase transformation from the trigonal to the monoclinic system.*

| Wyckoff-site (Trigonal symmetry) | Wyckoff-site (Monoclinic symmetry) |
|---|---|
| 3a | 2a |
| 3b | 2c |
| 6c | 4i |
| 18h | 4i |
|  | 8j |